\documentclass{article}
\usepackage{spconf,amsmath,graphicx}

\usepackage{hyperref}
\usepackage[ruled,linesnumbered]{algorithm2e}
\usepackage{amsfonts}
\usepackage{amsmath}

\DeclareMathOperator*{\argmin}{arg\,min}
\DeclareSymbolFont{CMsymbols}{OMS}{cmsy}{m}{n}
\SetSymbolFont{CMsymbols}{bold}{OMS}{cmsy}{b}{n}
\DeclareMathSymbol{\forall}{\mathord}{CMsymbols}{"38}
\DeclareMathSymbol{\exists}{\mathord}{CMsymbols}{"39}

\usepackage{multirow}
\usepackage{graphicx}
\graphicspath{ {./figures/} }
\usepackage{caption}
\usepackage{subcaption}
\usepackage[bottom]{footmisc}
\usepackage{yonatan}
\usepackage{cite}

\title{Optimizing Wi-Fi Channel Selection in a Dense Neighborhood.\\A Technical Paper prepared for SCTE Technical Journal}
\name{Yonatan Vaizman, Hongcheng Wang}
\address{Comcast --- Applied AI \& Discovery}
\begin{document}
%
\maketitle
\begin{abstract}
In dense neighborhoods, there are often dozens of homes in close proximity. This can either be a tight city-block with many single-family homes (SFHs), or a multiple dwelling units (MDU) complex (such as a big apartment building or condominium). Each home in such a neighborhood (either a SFH or a single unit in a MDU complex) has its own Wi-Fi access point (AP). Because there are few (typically 2 or 3) non-overlapping radio channels for Wi-Fi, neighboring homes may find themselves sharing a channel and competing over airtime, which may cause bad experience of slow internet (long latency, buffering while streaming movies~\etc). Existing APs sometimes have smart channel selection features --- typically scanning the air to select the least occupied channel. However, because they work independently (the APs do not coordinate), this can cause a cascade of neighboring APs constantly switching channels, which is disruptive to the connectivity of the homes. Wi-Fi optimization over all the APs in a dense neighborhood is highly desired to provide the best user experience. 

We present a method for Wi-Fi channel selection in a centralized way for all the APs in a dense neighborhood. We describe how to use recent observations to estimate the potential-pain matrix: for each pair of APs, how much Wi-Fi-pain would they cause each other if they were on the same channel. We formulate an optimization problem --- finding a channel allocation (which channel each home should use) that minimizes the total Wi-Fi-pain in the neighborhood. We design an optimization algorithm that uses gradient descent over a neural network to solve the optimization problem. We describe initial results from offline experiments comparing our optimization solver to an off-the-shelf mixed-integer-programming solver. In our experiments we show that the off-the-shelf solver manages to find a better (lower total pain) solution on the train data (from the recent days), but our neural-network solver generalizes better --- it finds a solution that achieves lower total pain for the test data (``tomorrow'').

We discussed this work in the 2022 Fall Technical Forum as part of SCTE Cable-Tec Expo.
This paper was published in~\cite{wifioptscte}. For citing this work, please cite the original publication~\cite{wifioptscte}.~\footnote{See published paper at \url{https://wagtail-prod-storage.s3.amazonaws.com/documents/SCTE_Technical_Journal_V2N3.pdf}.}
\end{abstract}
\section{Introduction}
\label{sec:intro}
In dense neighborhoods, there are often dozens of homes in close proximity. This can either be a tight city-block with many single-family homes (SFHs), or a multiple dwelling units (MDU) complex (such as a big apartment building or condominium). Each home in such a neighborhood (either a SFH or a single unit in a MDU complex) has its own Wi-Fi access point (AP). Because there are few (typically 2 or 3) non-overlapping radio channels for Wi-Fi, neighboring homes may find themselves sharing a channel and competing over airtime, which may cause bad experience of slow internet (long latency, buffering while streaming movies~\etc). Existing APs sometimes have smart channel selection features --- typically scanning the air to select the least occupied channel. However, because they work independently (the APs do not coordinate), this can cause a cascade of neighboring APs constantly switching channels, which is disruptive to the connectivity of the homes. Wi-Fi optimization over all the APs in a dense neighborhood is highly desired to provide the best user experience. 

We present a method for Wi-Fi channel selection in a centralized way for all the APs in a dense neighborhood. We describe how to use recent observations to estimate the potential-pain matrix: for each pair of APs, how much Wi-Fi-pain would they cause each other if they were on the same channel. We formulate an optimization problem --- finding a channel allocation (which channel each home should use) that minimizes the total Wi-Fi-pain in the neighborhood. We design an optimization algorithm that uses gradient descent over a neural network to solve the optimization problem. We describe initial results from offline experiments comparing our optimization solver to an off-the-shelf mixed-integer-programming solver. In our experiments we show that the off-the-shelf solver manages to find a better (lower total pain) solution on the train data (from the recent days), but our neural-network solver generalizes better --- it finds a solution that achieves lower total pain for the test data (``tomorrow'').

We discussed this work in the 2022 Fall Technical Forum as part of SCTE Cable-Tec Expo.
This paper was published in~\cite{wifioptscte}. For citing this work, please cite the original publication~\cite{wifioptscte}.

\section{Wi-Fi Pain Metric}
\label{sec:wifipain}
To measure the pain caused to the users in a dense Wi-Fi space, we define a new Wi-Fi pain metric. The main cause for Wi-Fi density pain is when a home's neighbors are using the same radio channel and occupying much of its airtime: when my home's AP senses high interference because others are using the same channel, my home's devices (including my AP) will have to wait longer times before they can send their packets over the radio channel, and this will cause slowness and subpar user experiences.

However, if my home barely has internet traffic during the night, while my neighbors use the same Wi-Fi channel heavily at the same time, that interference doesn't cause me any pain. The pain comes when my neighbors use the channel heavily while my home tries to use the same channel.

In addition, my home may have a lot of internet traffic at the same time as another home in my apartment building, but because there are five floors separating the two homes, our Wi-Fi signals never interfere with each other (the homes cannot ``sense'' each other --- we will define this more formally later).

To simplify, we notice that in a dense neighborhood, homes cause each other Wi-Fi pain when three conditions are met: the homes can sense each other, they tend to have a lot of internet traffic at the same times, and they use the same radio channel. The first two are regarded as given conditions of the neighborhood (we can measure or estimate them, but we cannot control them) and the third is the aspect that we can control --- which channel does each home use. We treat these three components as independent. Let's now formalize the overall pain mathematically with these three components, for a neighborhood with $n$ homes and $n_c$ Wi-Fi channels:

\begin{itemize}
\item The \textbf{(binary) sensing matrix}, $S^b\in\{0,1\}^{n\times n}$. $S^b_{i,j}$ is $1$ if and only if home $i$ can sense (and be interfered by) home $j$.
\item The \textbf{co-usage matrix} $U\in\mathbb{R}_+^{n\times n}$. This describes how much homes tend to have internet traffic at the same time. Notice, it doesn't matter which channel each home is using, and it doesn't matter if the homes can sense each other. This component only cares about the behavior patterns of the homes' residents and devices (specifically, the internet-activity patterns).
\item The \textbf{channel allocation matrix}: $C\in\{0,1\}^{n\times n_c}$. For each home (row), which channel is assigned to it --- exactly one channel (out of the $n_c$ options) has a value of $1$. Typically, $n_c$ is $2$ or $3$.
\end{itemize}

The pain that home $j$ causes to home $i$ depends on the three conditions we mentioned --- this is expressed with multiplication:
\begin{equation*}
\sum_{c=1}^{n_c}{ S^b_{i,j} U_{i,j} C_{i,c} C_{j,c} }.
\end{equation*}

Notice, that we use matrix $C$ twice in the formula and inside a sum over the possible channels ($c$) --- this is to capture if the two homes are using the same channel: if the two homes are not on the same channel, the whole sum will be $0$, but if they are on the same channel, the sum will have a single non-zero element $S^b_{i,j} U_{i,j}$. Similarly, if the two homes don't even sense each other ($S^b_{i,j}=0$), the whole sum will be $0$ (even if they are using the same channel) --- this can describe two homes that are physically far away from each other in the neighborhood, or have many walls between them, so the radio signal doesn't travel from one to the other. We assume additivity: the pain that home $i$ senses from the neighborhood is the sum of the pain that it senses from all the neighborhood's homes:
\begin{equation*}
pain_i = \sum_{c=1}^{n_c} \sum_{j=1}^n {{ S^b_{i,j} U_{i,j} C_{i,c} C_{j,c} }}.
\end{equation*}

To simplify the formula, we combine the two components that we cannot control and define the potential-pain matrix $P=S^b \odot U$ (elementwise multiplication). $P_{i,j} = S^b_{i,j} U_{i,j}$ describes the pain that home $j$ would add to home $i$ if they were using the same channel. The total pain in the neighborhood is a sum over the homes:

\begin{equation*}
pain^{total} = \sum_{c=1}^{n_c} \sum_{i,j=1}^n {{ S^b_{i,j} U_{i,j} C_{i,c} C_{j,c} }} = \sum_{c=1}^{n_c} \sum_{i,j=1}^n {{ P_{i,j} C_{i,c} C_{j,c} }}
\end{equation*}
And we can express it in matrix form:
\begin{equation*}
pain^{total} = \sum_{c=1}^{n_c} { \left[ C^TPC \right]_{c,c} } = Tr(C^TPC)
\end{equation*}

\begin{figure}
\centering
\begin{subfigure}[t]{0.95\linewidth}
\includegraphics[width=\textwidth]{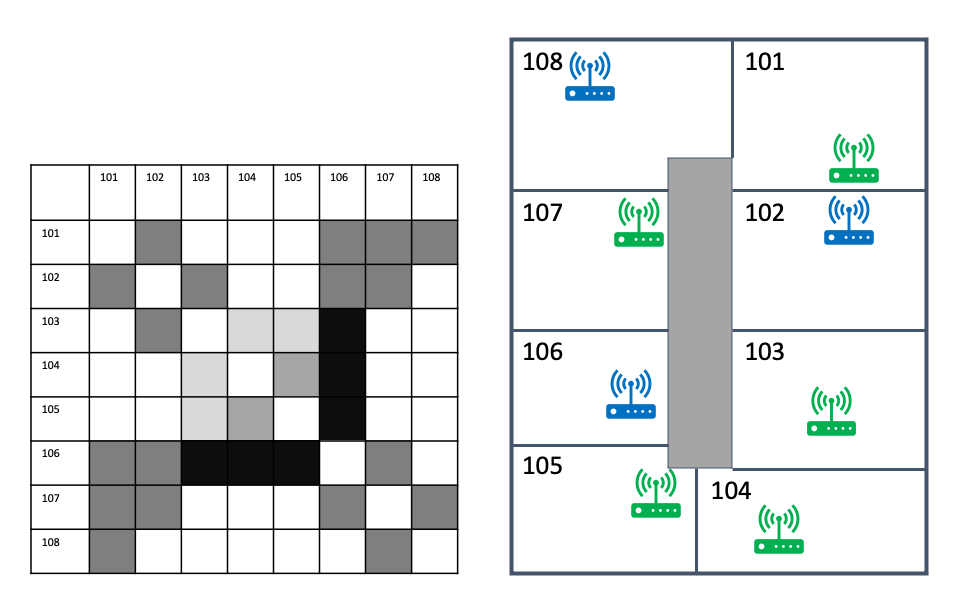}
\caption{Channel Allocation for a Dense Area}
\label{fig:dense_area}
\end{subfigure}
\end{figure}

Figure~\ref{fig:dense_area} illustrates part of a made-up dense neighborhood (right image) --- a floor plan with 8 apartments in an apartment building, and the potential-pain matrix for the 8 homes (left image), where darker shades of gray represent higher potential-pain value. The floor plan in the figure has two colors to the APs in the homes, representing a possible channel-allocation to two channels (blue and green).

Homes 101 and 104 are far away from each other (see the floor plan), so their APs never sense each other --- this explains why they have a blank (0) value in the matrix --- they have 0 potential to cause each other pain. This also explains why a smart channel allocation may allocate the same channel (green) to these two homes.

Home 106 represents a heavy internet user (most of the day has a lot of traffic), so it has the potential to cause much pain (darker shade in the matrix) to the homes that can sense it and typically have internet traffic at the same times (103, 104, 105). Homes 101 and 102 can sense home 106, but they may have internet traffic at different times of the day than home 106, so they have lower potential pain from 106 (medium gray shade). It makes sense to put home 106 on the blue channel and isolate it from homes 103, 104, and 105 (allocated the green channel).

\section{3.	Optimization Problem and Solvers}
\label{sec:optimization}
We can now define the main optimization problem as follows:
\begin{align}
C^* &= \argmin_{C\in\{0,1\}^{n\times n_c} }{ Tr(C^TPC) } \\
      &s.t. \\
      &\forall i \in \{1 \ldots n\}: \sum_{c=1}^{n_c}{C_{i,c}} = 1
\end{align}

This problem assumes we know (or estimate from recent data) the potential-pain matrix P --- it is the conditions of the neighborhood, the potential of homes to cause Wi-Fi pain to one another. The task of the optimization is to select a good combination of per-home channels, to minimize the overall pain that the homes cause each other. One of the reasons for this centralized channel selection approach is to avoid too many channel changes --- frequent changes can be disruptive to the users' connectivity experience. So, a typical use would be to solve this optimization problem, set the selected channels to all the neighborhood’s APs, and keep the channels fixed for a while (\eg~a whole day, a whole week).

\subsection{MIQP Problem Solver}
\label{ssec:miqp}
We note that our optimization problem is a mixed-integer quadratic programming (MIQP) problem: the search parameter $C$ appears in the objective function (the formula for total pain) in a quadratic form, and its values are constrained to be integers. This is a non-convex problem, and we don’t have an algorithm that can guarantee finding the global optimum (the very best combination of per-home channels) in reasonable time.

There are commercially available solvers, like Gurobi~\cite{gurobi}, that use a branch-and-bound approach to solve mixed integer programming problems (including the quadratic type). These methods iteratively try to rule out parts of the parameter-space and narrow down where we can find the global optimum, as well as narrow down the gap between lower and upper bounds for the optimal objective value. These tools often manage to reach the global optimum and they employ various heuristics to try to speed up the process.

\subsection{Neural Network Gradient Descent}
\label{ssec:neuralnet}

\begin{figure}
\centering
\begin{subfigure}[t]{0.95\linewidth}
\includegraphics[width=\textwidth]{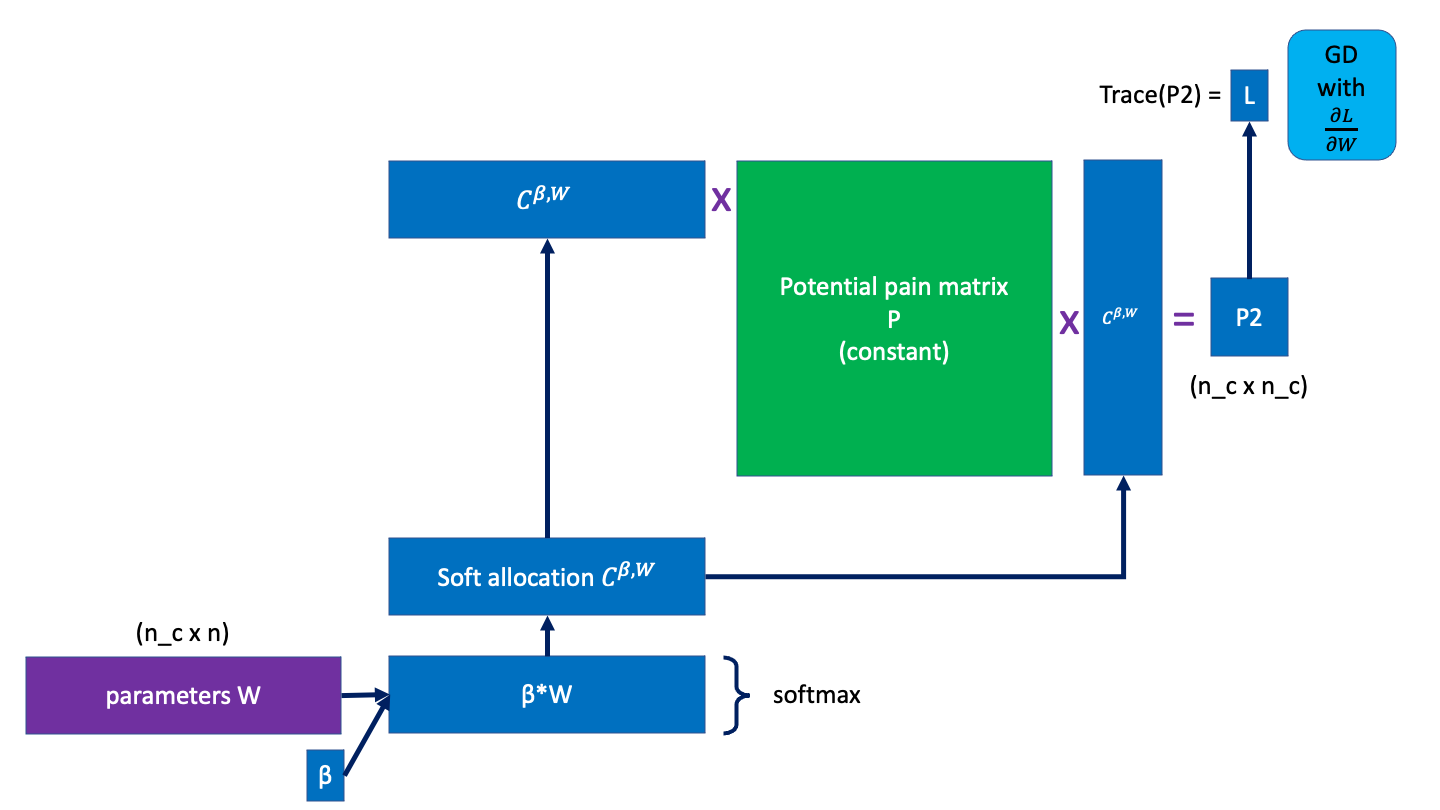}
\caption{Neural Network solver with gradient descent.}
\label{fig:neural_network}
\end{subfigure}
\end{figure}

We propose an alternative method to solve the optimization problem. We construct a neural network model to calculate a soft-approximation of the neighborhood's total pain, given any combination of channel allocation, and use gradient descent with back-propagation to change the underlying parameters until the pain reduces to a local minimum. The model is illustrated in figure~\ref{fig:neural_network}.

The model's parameters are represented as a matrix $W\in \mathbb{R}^{n\times n_c }$. The input to the model is a dummy scalar variable $\beta \in \mathbb{R+}$. Using $W$ and $\beta$, the model calculates a ``soft'' version of channel allocation $C^{\beta, W} \in \left[0,1\right]^{n \times n_c}$ by using the softmax operation on each row of $\beta W$: 
\begin{equation}
C^{\beta, W}_{i,c} = \frac{ e^{\beta W_{i,c}} }{ \sum_{d=1}^{n_c}{ e^{\beta W_{i,d}} } }
\end{equation}

The resulting matrix $C^{\beta, W}$ has each row (for home $i$) describing a probability distribution over the $n_c$ optional channels. This is not a valid channel allocation (in practice each AP only uses a single channel at a time), but this is a soft approximation of a valid channel allocation.
The model then incorporates the potential pain matrix $P$ as a fixed given input and uses $C^{\beta, W}$ to calculate a soft approximation of the total pain:
\begin{equation}
pain^{\beta, W} = Tr({C^{\beta, W}}^TPC^{\beta, W})
\end{equation}

Notice, that the input variable $\beta$ controls the order of the approximation: with a small value, like $\beta=0.1$ the soft channel allocations in $C^{\beta, W}$ will be closer to a uniform distribution over the $n_c$ channels. With a higher value, like $\beta=100$, the soft channel allocations better approximate a valid channel allocation --- where for each home only a single channel gets a value close to $1$ and the other channels get a value close to $0$.

To solve the optimization problem, we start by randomly initializing the parameters $W$ (\eg~using an \emph{i.i.d.} standard normal distribution), and then use gradient descent (with back-propagation) to reduce the approximated total pain $pain^{\beta, W}$. In addition, we start by using a small value of $\beta$ as input, and slowly increase it. This helps the algorithm first find a good global area and only later fine tune the parameters to a local minimum. After this procedure converges to a local minimum, and the parameters are tuned to values $W^{end}$, we can get the solution (the chosen channel allocation) by looking at the approximated channel allocations (for large $\beta$) and thresholding their values: 
\begin{equation}
C_{i,c}^{end} = 1\left[ C_{i,c}^{1000,W^{end}} > 0.5 \right].
\end{equation}

While this gradient descent approach does not presume to find a better (lower) optimum than off-the-shelf solvers, we want to highlight a few advantages it has:
\begin{itemize}
\item This approach doesn't assume that the potential pain matrix $P$ is symmetric, while other methods may rely on convex relaxations of the optimization problem, requiring them to have a symmetric (and positive-semi-definite) matrix for the quadratic form.
\item This approach can be modified to solve a different optimization problem that tries to minimize the worst-home-pain instead of the total, or average-home-pain. By making slight changes to the neural network, it can approximate the pain of the worst suffering home, and the optimization will try to minimize that value.
\item It is simple to add modifications that are common in training neural networks for typical supervised machine learning. We can use momentum when updating the parameters, for faster convergence. We can add parameter regularization (like the $L_2$ norm penalty --- $\lambda ||W||_2^2$) to the loss function, to avoid ``overoptimizing'' to the estimated potential-pain matrix.
\item This approach runs efficiently, quickly reaching a local minimum.
\item This approach does not overemphasize getting to the global minimum. We want to generalize to near-future data, so we should avoid overfitting to the most recent days' data.
\end{itemize}

\section{Preliminary Experiments}
\label{sec:experiments}
During summer 2021, we conducted a few offline experiments with data from a big apartment building. We had data from 66 homes in the building, so we treated them as ``the neighborhood's homes'' for the experiment. We tried various combinations of different aspects, and we share here some of our preliminary experiments and results. In these experiments, we simulated running the optimization on a reference date, to select the channel allocation for the following day. We collected data from the homes in the neighborhood from the recent days up to (and including) the reference day (the train days), calculated the potential pain matrix, and solved the channel allocation problem. We did a similar calculation to get the potential pain matrix for the day following the reference day (the test day). We evaluated the total pain on both the train days and the test day, given the chosen channel allocation, keeping in mind that the real goal is to improve (minimize) the pain on the test day.

\subsection{Estimating Potential Pain}
\label{ssec:estimating}

\begin{figure}
\centering
\begin{subfigure}[t]{0.95\linewidth}
\includegraphics[width=\textwidth]{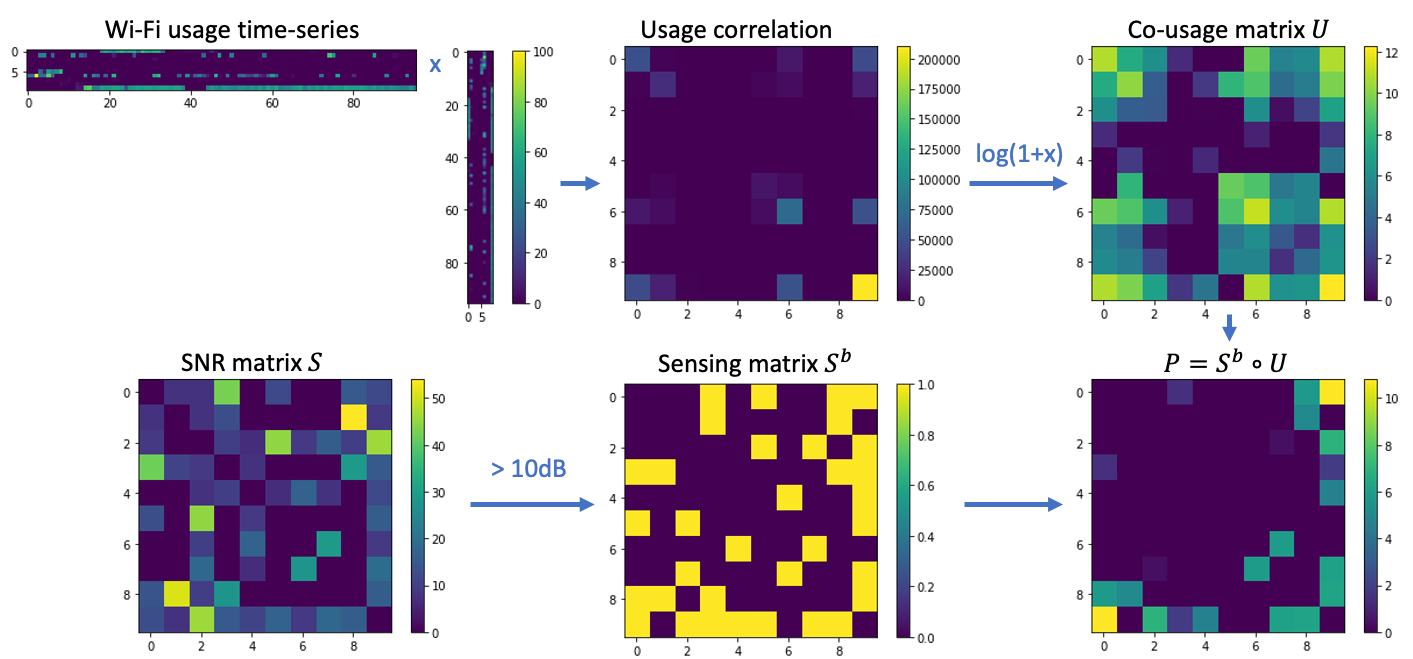}
\caption{Estimating the Co-Usage, Sensing, and Potential Pain Matrices.}
\label{fig:estimating_p}
\end{subfigure}
\end{figure}

We estimated the two components of the potential pain separately: the (binary) sensing matrix $S^b$ and the internet co-usage matrix $U$. Figure~\ref{fig:estimating_p} illustrates this process: the colors in the matrices represent the cell values, ranging from 0 (dark blue) to high values (bright yellow). Each matrix has a different range (see the color-bar to the right of each image).

The co-usage can be defined as some version of multiplying two home's internet-traffic time-series ($u_{t,i}$ represents home $i$'s usage at time $t$). In this paper we use $U_{i,j}=\log(1+ \sum_t{u_{t,i}u_{t,j}})$, but we can have many variations: sum each home’s time-series first and then multiply, use a different non-linearity than logarithmic, apply non-linearity on $u_{t,i}$ alone to produce a non-symmetric version,~\etc\ To estimate the co-usage matrix U, we used periodic measurements that each AP takes every 15 minutes. Specifically, we used a measurement of percentage of airtime that the AP occupied the channel to transmit data to the home's devices (the ``download'' direction, assumed to occupy the majority of airtime in a typical home). We smoothed the quarter-hourly measurements to hourly quantities. We experimented with both measurements from whole-days (all hours of the day) and evening-time (only using measurements from 7pm-10pm local time), but here we focus our results on evening-time. 
For estimation with the recent $n_d$ days, this results in a time-series (vector) of $3n_d$ hourly values for each home. We then calculated the cross-correlation between homes (the dot product of two homes' time-series) and took the $\log(1+x)$ of these values.

The top row of figure~\ref{fig:estimating_p} illustrates the process of estimating the co-usage matrix: starting with a Wi-Fi usage time-series for each home (top left). The image shows 10 homes and airtime-percentage values from 96 time points. This narrow matrix is multiplied by its transpose to produce the usage correlation matrix (for each pair of homes the value is the dot product of their two time-series). These correlation values can be extremely large (notice the color-bar reaching values of 200k), so we then apply logarithmic compression to form the co-usage matrix $U$.

For estimating the sensing matrix $S^b$, we used radio-scan reports from the APs in the neighborhood: each AP performs a scan multiple times a day to look for Wi-Fi beacons in the air. The AP records the media access control (MAC) address of every other AP that it senses, and the signal to noise ratio (SNR) of the sensed beacon. We mapped sensed Wi-Fi MAC addresses to the familiar APs that are part of the neighborhood. The scans reported additional sensed entities that came from external APs (which we don't know and cannot control). For each pair of homes $<i,j>$ in the neighborhood, we averaged the SNR values (over a period, like a week) of how strongly home $i$'s AP senses home $j$'s AP. We can call these variables the SNR matrix $S$ (typically having non-negative real values), illustrated in figure~\ref{fig:estimating_p} bottom left image for 10 homes. In our experiments, we chose to symmetrize the sensing matrix: $S \leftarrow 0.5(S+S^T)$. We applied a threshold of 10dB to produce the binary sensing matrix $S^b$ (figure~\ref{fig:estimating_p}, bottom middle image). Notice that since an AP never sensed itself in the radio scans, we naturally get zeros in the diagonal. This fits our formulation, because we wish to only model the pain that homes cause other homes, not themselves.

We multiplied (elementwise) these two estimated matrices $U$ and $S^b$ to form the potential pain matrix $P$ (bottom right image in figure~\ref{fig:estimating_p}). 

For the test day, we calculated the co-usage matrix $U$ from the test day's usage measurements. However, we used the same SNR matrix $S$ as we did for the train days. This is because we didn’t have sufficient scan measurements from every day, and because we assumed that who can sense whom stayed stationary over a longer time (approximately a month).

\subsection{Optimization Details}
\label{ssec:opt_details}
We used the Gurobi package~\cite{gurobi} as a MIQP solver. For our neural network algorithm, we implemented the network using TensorFlow~\cite{tensorflow} and Keras~\cite{keras}. Every update step had just a single example input into the network. We increased the value of the input variable $\beta$ in phases (running 6,400 update steps in each phase) with values: 1, 10, 100, 1000. We used ADAM optimizer with learning rate 0.001.

\subsection{Results}
\label{ssec:results}

\begin{center}
\begin{tabular}{ |c|c|c|c| }
\textbf{Train days} & \textbf{Algorithm} & \textbf{Train pain} & \textbf{Test pain} \\
 1 (Aug 24) & Gurobi & 58.2 & 194.0 \\ 
 1 (Aug 24) & Neural Network & 58.2 & 184.9 \\ 
 4 (Aug 21--24) & Gurobi & 64.5 & 166.3 \\ 
 4 (Aug 21--24) & Neural Network & 69.9 & 143.8 \\ 
\end{tabular}
\label{tab:results}
\end{center}

We show in table~\ref{tab:results} results from a few of our offline experiments with a single neighborhood. These were all done with train days up to (and including) August 24th and testing on usage data from August 25th. In these experiments, we used usage (and scan information) from the 2.4GHz frequency and we simulated solving the channel allocation for $n_c=2$ channels. Rows 1--2 show experiments where there was only a single training day, compared to 4 training days in rows 3--4 (the table reports the average total pain per train day). The results show that when training with data from more days, we could achieve a worse (higher) total pain on the train data, but a better (lower) total pain on the test day, which is what we want to achieve. As expected, our neural network solver did not beat Gurobi's solution on the train days. However, the neural network solver's solution generalized better to the test day --- it achieved a lower pain than Gurobi's solution (in both the 1-train-day and 4-train-days scenarios).

\section{Conclusions}
\label{sec:conclusions}
We have discussed the problem of Wi-Fi airtime competition in a dense neighborhood and the need for a centralized channel selection solution. We defined a Wi-Fi pain objective, based on the co-occurrence of close neighbors having a lot of internet traffic at the same time on the same radio channel. We formulated the pain such that all the relevant information is captured in a single square matrix $P$, indicating for each pair of homes how much pain would one add to the other if they were using the same channel. We formulated an optimization problem and offered two alternative solvers for it: an off-the-shelf MIQP problem solver and a tailored neural network solver. We conducted preliminary offline experiments with data from a real neighborhood and demonstrated how we can achieve better generalization (lower pain for ``tomorrow'') with more training days and by using our neural network solver.

\subsection{Future Directions}
\label{ssec:future}
There are still many more directions to research. We can explore various flavors of defining Wi-Fi pain: there can be non-symmetric definitions of potential-pain, for example, when one AP tends to transmit with higher power than a neighboring AP. We can incorporate external sources of interference into the pain model, for example it is possible that the lower floors of an apartment building consistently experience interference in a particular radio channel from a nearby store. When estimating the potential-pain based on the recent month, we may want to give different weight to different days of the week, to get a better estimation of what is about to happen tomorrow. Different neighborhoods may require different approaches --- a suburban condominium with long term residents may be a good candidate for estimating the potential-pain based on a whole month, while a big apartment building in a busy city block may have faster turnaround of residents and may require estimation based on the most recent few days.

The optimization algorithm can have various adjustments. We can add regularization on the parameters W, or even constraints on the output values of some of the nodes in the neural network. The schedule of changing $\beta$ may influence the outcome. An interesting direction is minimizing the worst-home pain and seeing how it influences the average-home pain. To explore this direction, we need to adjust the neural network: instead of calculating the whole quadratic form of $C^{\beta,W}$, the network will first calculate the approximated total pain for each home individually, and then apply a soft approximation of the max operation, to pick the most suffering home in the neighborhood. 

We will conduct more offline experiments with many more neighborhoods. Additionally, actual trials will reveal more reliably how helpful is centralized channel selection and A/B tests can help demonstrate which methods are better. We can use a contextual-bandit approach to cleverly select the appropriate ``flavor'' of Wi-Fi pain for each neighborhood (\eg~how many days to use when estimating the potential-pain). In actual channel-selection experiments, we can more directly measure the sensed interference that every AP experiences from its environment. More importantly, we’ll have to assess the effect on the residents’ subjective experience of slow internet and Wi-Fi pain.

\section{Acknowledgements}
We thank Zhehao (Kenny) Zhang for running the offline experiments, and Shelley Leung for processing telemetry data. We thank Obi Asinugo, Zekun (Katherine) Yang, Ali Mohammadi, Teddy ElRashidy, and Colleen Szymanik for great discussions and guidance about this interesting Wi-Fi problem.

%


\vfill\pagebreak


\bibliographystyle{IEEEbib}
\bibliography{refs}

\end{document}